# Is Feynman's Analysis of Electrostatic Screening Correct?

Kris Subbarao[1]
Sukhbir Mahajan[2]

## Abstract

In a recent paper on electrostatic shielding, Chapman, Hewett and Trefethen present arguments that the analysis of this effect by Feynman is incorrect. Feynman analyzed shielding by a row of infinitesimally thin wires. They claim that Feynman used the wrong boundary condition invalidating his analysis. In this paper we emphasize that Feynman's solution is a Green's function through which behavior of the potential due to finite thickness wires of arbitrary cross-section with appropriate boundary condition can be understood. This shows that the main conclusions of Feynman's treatment are indeed correct. The configuration analyzed by Maxwell of parallel plates with one of the plates replaced by a row of wires provides a more intuitive understanding of Feynman's argument. The case of a *finite* number of wires arranged in a ring (not treated by Feynman), as model of Faraday cage, is different. The structure of the solution in radial coordinates already suggests that one should not expect exponential behavior and further analysis confirms this. There appear to be contradictory results in the literature for its screening behavior in the presence of external charges/fields.

[1] KrisSubbarao at gmail.com
[2] Mahajans at csus.edu

## 1. INTRODUCTION

In a recent article on Faraday cage effect, Chapman, Hewett and Trefethen [1] state that "One of the few mathematical treatments we have found is in section 7-5 of Vol. 2 of The Feynman Lectures on Physics, where so far as we can tell, the analysis is incorrect". The Feynman reference is to [2]. In a related article [3], it is stated that "Nick Trefethen challenges Richard Feynman's well-known analysis of the Faraday cage effect and exponential shielding", and "The error is that Feynman's wires have constant *charge*, not constant *voltage*. It's the wrong boundary condition! I think that Feynman, like me and most others beginning to think about this problem, must have assumed that the wires may be taken to have zero radius. The trouble is, a point charge makes sense, but a point voltage does not. (Dirichlet boundary conditions for the Laplace equation can only be applied on sets of positive capacity.) Since the correct boundary condition cannot be applied at points, I'm guessing Feynman reached for one that could, intuiting that it would still catch the essence of the matter. This is a plausible intuition, but it's wrong."

It is the purpose of this article to show that Feynman's analysis is actually correct; the infinitesimal radius of the wires is *not* a problem. The authors, however, correctly point out that a complete treatment of electrostatic shielding should include realistic external fields and charges and how they induce charges on the conductors.

Electrostatic shielding itself is a subset of electromagnetic shielding. A Faraday cage should shield against time varying electromagnetic fields, should take into account polarization of the waves, skin effect etc. Electrostatic shielding is important in many circumstances and is being considered as a method to shield humans in spacecraft especially for deep space explorations.

Maxwell [4] studied the electrostatic problem, in particular, of a row of wires (also called a grating) in between two parallel plates under various conditions; such studies are now textbook material. In his lectures, Feynman discussed a subset of problems considered by Maxwell. Many details are given in Maxwell's treatise.

## 2. PRELIMINARIES

Let us do a quick review of standard textbook material. Starting with the potential of a point charge $e/\varepsilon_0$ at the origin, $\Phi_0(r) = \frac{e}{4\pi|\mathbf{r}|}$, let us define a free space Green's function (sometimes referred to as impulse response):

$$G_1(\mathbf{r}, \mathbf{r}') \stackrel{\text{def}}{=} \frac{1}{|\mathbf{r}-\mathbf{r}'|}, \qquad (1)$$

obeying the equation $\nabla^2 G_1(\mathbf{r}, \mathbf{r}') = -4\pi\delta(\mathbf{r} - \mathbf{r}')$. For simplicity we will drop $\varepsilon_0$ from this point on.

For a *known* charge distribution $\rho(\mathbf{r}')$, localized in a volume V, the potential $\Phi(\mathbf{r})$ is given by

$$\Phi(\mathbf{r}) = \frac{1}{4\pi} \int \rho(\mathbf{r}') G_1(\mathbf{r}, \mathbf{r}') dV \qquad (2)$$

The integration is done as **r'** varies over the volume *V*. If the charge distribution is not known a priori, but only boundary values of the potential (e.g., Dirichlet conditions) in certain regions of V, it is often expedient to define a different Green's function that differs by a harmonic function. The solution then can be used to determine the charge distribution. For points outside the region V, we can do a multipole expansion to represent the potential as a series in the powers of $\frac{1}{|r|}$ in the form:

$$\Phi(r) = \frac{\frac{1}{4\pi}\int \rho(r')dV}{|r|} + \frac{A_2}{|r|^2} + \cdots \quad (3)$$

The first term depends only on the total charge; the rest of the coefficients depend on the specifics of charge distribution.

Consider the special case of a charged conductor (satisfying Dirichlet boundary condition of constant potential on the surface) with no external charges. The potential satisfies the following equation:

$$\Phi(\mathbf{r}) = \frac{1}{4\pi}\oint \sigma(\mathbf{r}')G_1(\mathbf{r},\mathbf{r}')\,dA' \quad (4)$$

where $\sigma(\mathbf{r}') = -\frac{\partial \Phi(\mathbf{r}')}{\partial n'}$ is the surface charge density. An external field contributes an additional term to this equation. For example, for a constant external field E in the x-direction, it is $-Ex$.

The main lesson is that the asymptotic behavior of a well-defined potential satisfying well-defined boundary conditions can be analyzed through a free space Green's function. (Solving for the potential is often facilitated by another Green's function that incorporates the boundary condition. This need not concern us here since our main focus is the behavior of the solution, not the method of solution).

Consider now a line of charge. Although Green's function from Eq. ( 1 ) can be used for this case, it is expedient to define another Green's function that takes advantage of the symmetries of this system. Starting with the well-known potential for a line of charge of density $\lambda$ per unit length, $-2\lambda \log(|\mathbf{r}|)$, we can define a free space Green's function:

$$G_2(\mathbf{r},\mathbf{r}') \stackrel{\text{def}}{=} -\frac{1}{2\pi}\log(|\mathbf{r}-\mathbf{r}'|) \quad (5)$$

If we have an infinite cylinder of arbitrary cross section (not necessarily circular) with a charge distribution $\rho(\mathbf{r}')$ per unit area per unit length (where $\mathbf{r}'$ is a two-dimensional vector in a plane perpendicular to the cylinder). The potential is now given by

$$\Phi(r) = \frac{1}{4\pi}\int \rho(\mathbf{r}')\,G_2(\mathbf{r},\mathbf{r}')dA \quad (6)$$

The variable **r'** varies over the cross-sectional area A. In the limit of a wire of infinitesimal thickness with a charge $\rho(\mathbf{r}') = \lambda\,\delta(\mathbf{r}')$ with $\lambda$ as charge per unit length, we recover the potential for a line charge. (The presence of a logarithm with no scale (and one that diverges as $\mathbf{r} \to 0$) may cause some conceptual but no real issues. A modern renormalization point of view is presented in [5].) A multipole-like expansion gives an expression of the form

$$\Phi(\mathbf{r}) = -\left(\frac{1}{2\pi}\int \rho(\mathbf{r}')\,dA\right)\log|\mathbf{r}| + \frac{A_1}{|\mathbf{r}|} + \cdots \quad (7)$$

As before, we see that the free space Green's function defined for an infinitesimal wire is useful for studying the behavior of potentials satisfying realistic boundary conditions.

## 3. ROW OF WIRES

Feynman [2] analyzes a row of infinitesimally thin wires. This was studied by Maxwell [4] and is also a textbook problem (see e.g., [6] p. 292 Eq. 7). It is a basic problem studied in multiwire particle detectors. Although Feynman only presented the Fourier series coefficients, the series can actually be summed to obtain an explicit solution. Consider a row of wires in the $y=0$ plane running parallel to the z-axis, spaced a distance $a$ apart. Let one of the wires be at $x=0$. The potential, for a charge density $\lambda$ per unit length, is given by

$$\Phi_0(x,y) = -\frac{1}{4\pi}\frac{\lambda}{a}y - \frac{1}{4\pi}\frac{\lambda}{2\pi}\log\left(1 - 2e^{-\frac{2\pi y}{a}}\cos\left(\frac{2\pi x}{a}\right) + e^{-\frac{4\pi y}{a}}\right) \quad (8)$$

We can now construct a Green's function that takes advantage of the symmetries of the system:

$$G_3(x,y,x',y') \stackrel{\text{def}}{=} -\frac{1}{a}y + \frac{1}{a}y' - \frac{1}{2\pi}\log\left(1 - 2e^{-\frac{2\pi(y-y')}{a}}\cos\left(\frac{2\pi(x-x')}{a}\right) + e^{-\frac{4\pi(y-y')}{a}}\right) \quad (9)$$

This satisfies the equation

$$\nabla^2 G_3(x,y,x',y') = -4\pi \sum_{n=-\infty}^{\infty} \delta(x-x'-na)\delta(y-y') \quad (10)$$

This is applicable to problems with the required periodicity. For wires that are not infinitesimal, but have a finite arbitrary cross-section with charge distribution $\rho(x',y')$ per unit area per unit length, the potential is given by an integral over the cross-sectional area $A$:

$$\Phi(x,y) = \frac{1}{4\pi}\int G_3(x,y,x',y')\rho(x',y')dx'dy' \quad (11)$$

Note that $x'$ and $y'$ are constrained by the cross-section of the charge cylinder(s), $x$ and $x'$ are constrained by periodicity $-\frac{a}{2} \leq x, x' \leq \frac{a}{2}$.

When (9) is substituted into Eq.(11), the first term corresponds to a uniform electric field, the second is just a constant potential. The remaining term, can be expanded with $e^{-\frac{2\pi y}{a}}$ as the small parameter to give an exponential decay series. The exponents are powers of $-2\pi y/a$. Only the coefficients of the series terms depend on the details of the charge distribution.

This shows that Feynman's argument is correct. One just needs to recognize Feynman's solution for a wire of infinitesimal thickness as a Green's function and then apply it to finite arbitrary cross-sections. Equations such as (1), (5) and (9) obtained for infinitesimal objects allow us to examine certain behavior of realistic potentials obeying relevant boundary conditions, without an explicit solution. In fact, they often aid in the construction of the solutions.

The above argument only shows that a periodic row of identical wires, of finite thickness and arbitrary cross-section with appropriate charge, placed in front of a charged plate screens a uniform field outside the region between the plate and the wires. We have only invoked a periodic row of identical charged cylinders; we did not explicitly require the cylinders to be conducting. A full argument requires examining the behavior in the presence of external charges. This induces charges on the conductors to maintain them all at the same potential. How well the effects of the external charges are screened is the topic of interest.

Let us explain how Feynman's argument motivates screening by conductors. Consider a conducting plate A with a charge density $\sigma$ per unit area. By inserting another plate B with charge density $\sigma$, we can shield the region above B from the charges on the plate A. To determine potentials, insert a third grounded plate C at potential zero. The charges and potentials are shown in the left part of Figure 1.

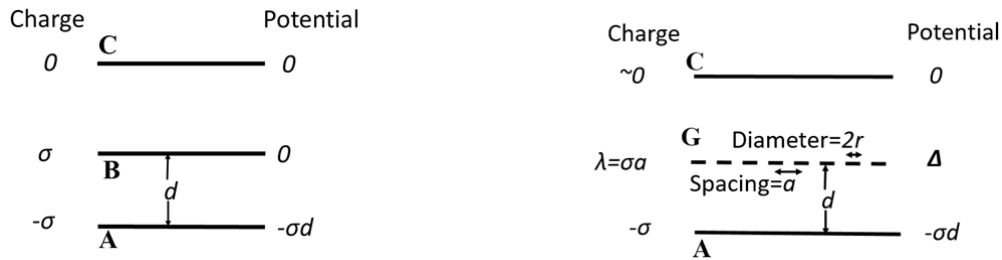

Figure 1: An assembly of three conducting plates. On the right the plate B is replaced by a row of wires (i.e., a grating)

Can we use, instead of the plate B, a grating G with wires of radius $r$ spaced a distance $a$ apart? To accomplish the same average charge density, the wires have to have a charge of $\sigma a$ per unit length. The field above G is shielded exponentially. This implies some tiny charges on the plate C. Note that the potential of the grating will be some value $\Delta$ depending on the spacing $a$ and the radius $r$ and $\Delta \to \infty$ as $r \to 0$. For any non-zero $r$, the potential $\Delta$ is finite, and the asymptotic behavior is given as described previously.

### 4. RING OF WIRES

Consider a ring of wires. Feynman did not consider this problem, but is of interest in connection with Faraday cage. It is well-known that in radial problems involving Laplace's equation, logarithm of the radial coordinate plays a natural role. This results in powers of the radius, in contrast to Cartesian problems resulting in exponentials. One should therefore expect power law screening in such geometries. Note that in the planar case, we start with an infinite number of wires; in the ring case, we start with a finite number of wires and study the limit when the number of wires is large.

Finite sized conductors in external fields/charges are perhaps best studied numerically. The authors of [1] should be commended for initiating such studies. In [1], wires of finite circular cross-section arranged uniformly in a circle is discussed in particular with external charges. First, let us consider some analytic results.

First consider N wires of infinitesimal cross-section arranged uniformly in a circle of radius $R$. This is the ring analog of the row of wires addressed by Feynman. The solution for the resulting potential is ( [6] p. 290)

$$\Phi_0(r,\theta) = \frac{1}{4\pi}\frac{\lambda}{2N}\log\left(1 - 2\left(\frac{r}{R}\right)^N \cos(N\theta) + \left(\frac{r}{R}\right)^{2N}\right) \tag{12}$$

As before, we can construct a Green's function:

$$G(r,\theta,r',\theta') = \frac{1}{2N}\log\left(1 - 2\left(\frac{r}{r'}\right)^N \cos(N(\theta - \theta')) + \left(\frac{r}{r'}\right)^{2N}\right) \qquad (13)$$

Because of periodicity, $-\frac{\pi}{N} \leq \theta, \theta' \leq \frac{\pi}{N}$. One can now construct the analog of Eq. (11) for wires of finite cross-section of arbitrary shape. For the region $r < R$, where R is a characteristic distance of a cross-section from the origin, we get an expansion in powers of $\left(\frac{r}{R}\right)^N$. This exhibits a type of screening, the leading term being of the form $A_1 \left(\frac{r}{R}\right)^N$. Larger the value of N, the more effective this screening.

As in the case of parallel plates, we can start with a uniformly charged cylinder, and examine screening by surrounding it by a ring of grounded conducting wires. Outside the ring of wires, the fields fall off as a power of the distance, with the power determined by the number of wires. A fuller examination requires introducing different external fields and charges. It is worth noting that for wires of finite radius, for large enough N, the wires will overlap.

Perhaps the next simplest problem is introducing a uniform external field *E*, let us say in the x-direction. Starting with a ring of *N* wires of radius *a*, and making various approximations including smallness of *a*, Martin [7] concludes that the potential is approximately given, for large *N* and *r* < *radius of the ring*, by

$$-2E\frac{\log N}{N} r \cos\theta$$

which results in screening even in the limit of wires of infinitesimal thickness; inside the ring, the electric field is in the same direction as the unperturbed field but only $2\frac{\log N}{N}$ times as strong. In contrast, in [1], the authors numerically studied a ring of wires of finite radius *a* in the presence of a single external (line of) charge and concluded (with a simple change for notational conformity) that "the Faraday shielding effect depends upon the wires having finite radius and is weaker than one might expect, scaling as | log *a*|/N in an appropriate regime of small *a* and large *N*." Although the geometry of the external charge/field in [7] is different from that in [1], the difference in the screening behavior appears to be contradictory.

### 5. CONCLUSIONS
1. Feynman's analysis of screening by a row of *infinite* number of infinitesimal wires is correct. One just has to recognize his solution as a Green's function from which solutions for wires of finite cross-section can be constructed. The asymptotic behavior is not dependent on the finiteness of the cross-section in much the same way as a finite distribution of charges asymptotically behaves as a single charge of value equal to the sum of the charges. The singularity for infinitesimal thickness of the wires is no more problematic than the singularity for a point charge.
2. The asymptotic behavior is exponential. The exponents depend only on the spacing between wires and not on the finiteness of the wires. The coefficients multiplying the exponential terms depend on the wire cross-section. In the well-defined limit of infinitesimal wires, one recovers Feynman's results. Analysis of parallel plates with one of the plates replaced by a row of wires, a configuration analyzed by Maxwell, provides a more intuitive understanding of Feynman's argument.

The case of a *finite* number of wires arranged in a ring, a model of Faraday cage, is different. (Feynman does not discuss this nor does he use the word "cage" or the phrase "Faraday cage" in the relevant section). The structure of the solution in radial coordinates already suggests that one should not expect exponential behavior and further analysis confirms this. There appear to be contradictory results in the literature for the screening behavior in the presence of external charges/fields.

Feynman's analysis is a simplified picture of electrostatic screening. Considerable additional analysis is needed for a practical theory and design of electromagnetic screening in realistic situations.

## 6. ACKNOWLEDGMENTS

We thank Professor Kirk McDonald of Princeton University for insightful comments.